\documentclass[runningheads]{llncs}

\usepackage{amssymb}
\usepackage{hyperref}
\usepackage{amsmath}
\usepackage{verbatim}
\usepackage[square,sort,comma,numbers]{natbib}
\usepackage[nohyphen]{underscore}
\usepackage[capitalise]{cleveref}
\usepackage{csquotes}
\usepackage{changebar}

\usepackage{booktabs}

\usepackage{ifpdf}

\usepackage{graphicx}

\usepackage{color} %

\usepackage{mathtools}

\usepackage{epstopdf}

\usepackage{listings}

\usepackage{scalerel}

\usepackage{tikz}
\usetikzlibrary{shapes,arrows,automata,positioning,cd}
\tikzset{
  cfgedge/.style   = {black, ->, >=stealth},
  forward/.style = { blue, ->, >=angle 45},
  backward/.style = { red, densely dashed, ->, >=latex' },
  backwardleft/.style = { red, densely dashed, <-, >=latex' },
}
\usepackage{xcolor}

\usetikzlibrary{fit,calc}

\colorlet{pink}{red!40}
\colorlet{cyan}{cyan!60}

\usepackage{prettyref}

\newrefformat{thm}{Theorem~\ref{#1}}
\newrefformat{lem}{Lemma~\ref{#1}}
\newrefformat{cha}{Chapter~\ref{#1}}
\newrefformat{sec}{Section~\ref{#1}}
\newrefformat{tab}{Table~\ref{#1}}
\newrefformat{fig}{Figure~\ref{#1}}
\newrefformat{alg}{Algorithm~\ref{#1}}
\newrefformat{exa}{Example~\ref{#1}}
\newrefformat{def}{Definition~\ref{#1}}
\newrefformat{li}{Line~\ref{#1}}

\usepackage{enumitem}
\newlist{ecomponents}{enumerate}{1}
\setlist[ecomponents,1]{label={\bfseries C\arabic*},align=left}
\newlist{rqs}{enumerate}{1}
\setlist[rqs,1]{label={\bfseries RQ\arabic*},align=left}

\usepackage[ruled,vlined,linesnumbered]{algorithm2e}

\definecolor{light-gray}{gray}{0.9}
\definecolor{light-pink}{rgb}{0.858, 0.188, 0.478}

\newcommand{\Omit}[1]{}

\newcommand{\bugswarm}[0]{\textsc{BugSwarm}\xspace}
\newcommand{\criticalreview}[0]{\textsc{CriticalReview}\xspace}

\begin{document}
\title{A Note About: Critical Review of BugSwarm for Fault Localization and Program Repair}
\titlerunning{A Note About: Critical Review of BugSwarm}

\author{David A. Tomassi \and
Cindy Rubio-Gonz\'alez} %
\institute{University of California, Davis USA\\
\email{\{datomassi,crubio\}@ucdavis.edu}}

\maketitle              %
\begin{abstract}
Datasets play an important role in the advancement of software tools
and facilitate their evaluation. 
\bugswarm~\citep{DBLP:conf/icse/DmeiriTWBLDVR19} is an infrastructure to
automatically create a large dataset of real-world reproducible
failures and fixes.
In this paper, we respond to
\citet{DBLP:journals/corr/abs-1905-09375}'s critical review of the
\bugswarm dataset, referred to in this paper as \criticalreview. We
replicate \criticalreview's study and find several incorrect claims
and assumptions about the \bugswarm dataset. We discuss these
incorrect claims and other contributions listed by \criticalreview.
Finally,  we discuss general misconceptions about \bugswarm, and our
vision for the use of the infrastructure and dataset.

\end{abstract}

\section{Introduction}
\label{sec:Introduction}

Datasets are imperative to the development and progression of software
tools, not only to facilitate a fair and unbiased evaluation of their
effectiveness, but also to inspire and enable the community to advance
the state of the art. There have been various influential datasets
developed in the Software Engineering community (e.g.,
\citep{DBLP:conf/icse/HutchinsFGO94, Just:2014:DDE:2610384.2628055,
Do:2005:SCE:1089922.1089928, Lu05bugbench:benchmarks, 1555866,
DBLP:conf/kbse/DallmeierZ07, DBLP:journals/tse/GouesHSBDFW15}).
Unfortunately, these datasets have required a substantial amount of
manual effort to be created, which makes it difficult to grow them.

Recently we developed \bugswarm
\citep{DBLP:conf/icse/DmeiriTWBLDVR19}, an infrastructure that
leverages continuous integration (CI) to automatically create a
dataset of \emph{reproducible} failures and fixes. \bugswarm comprises
an infrastructure, dataset, REST API, and website. The initial dataset
(version 1.0.0 and reported in \citep{DBLP:conf/icse/DmeiriTWBLDVR19})
consists of 3,091 pairs of failures and fixes (referred to as
\emph{artifacts}) mined from Java and Python projects. Because artifacts
mined from open-source software are bound to have different
characteristics (number of failing tests, failure reason, fix
location(s), patch size, etc.), we provide a REST API and website for
users to navigate and select the artifacts that fit the needs of their
tools. \bugswarm is under active development, currently allowing the
mining of failures and fixes that satisfy specific characteristics.

Parallel to the development of \bugswarm,
\citet{DBLP:journals/corr/abs-1905-09375} conducted a review of the
\bugswarm dataset (version 1.0.1) with respect to Automated Program
Repair (APR) and Fault Localization (FL). The authors stated
characteristics they consider necessary for artifacts to be used in
studies that evaluate the state of the art in APR and FL.
Additionally, the authors presented a high-level classification of
failures, and discussed the cost of using \bugswarm artifacts. In the
rest of this paper we refer to
\citep{DBLP:journals/corr/abs-1905-09375} as \criticalreview.

One of the purposes of datasets is to facilitate the evaluation of
software tools. Instead, \criticalreview uses the general
requirements/current limitations of the state-of-the-art APR tools to
\emph{evaluate} the \bugswarm dataset. While it is important that
datasets possess key characteristics (e.g., failures that are relevant
to the tools under evaluation), the existence of artifacts that do not
have desired characteristics does not hinder studies if users can
navigate and select artifacts relevant to their studies. Limiting a
dataset to only include problems that certain tools can handle would
be of no benefit to our community. Furthermore, the goal of the
\bugswarm dataset is to identify the kinds of problems found in real
software and the environment in which these problems occur,
and thus inspire the community to advance
the state of the art.

In addition to general misconceptions on datasets, \criticalreview
discredits the use of the \bugswarm dataset based on multiple
\emph{incorrect} observations. Specifically, \criticalreview makes a
false allegation against \bugswarm paper
\citep{DBLP:conf/icse/DmeiriTWBLDVR19}'s reported data, and presents
wrong results and conclusions led by misunderstandings of
\textsc{Travis-CI} terminology and Docker's architecture.

This paper discusses each of \criticalreview's incorrect claims, which
had already been communicated to the authors of \criticalreview upon
their request for feedback \emph{prior} to the archival of their
study. We also discuss the two other contributions of \criticalreview:
a GitHub repository to store the code and build logs of the \bugswarm
artifacts, and \criticalreview's own website to browse \bugswarm
artifacts, both of which duplicate information already available in
\bugswarm.

The rest of this paper presents a brief overview of \bugswarm in
\cref{sec:bugswarm}, and describes the methodology used by
\criticalreview in \cref{sec:critical-review}. We discuss the
incorrect findings reported by \criticalreview in \cref{sec:corrections},
and the rest of the contributions of \criticalreview in \cref{sec:other-contributions}.
Finally, we clarify some misconceptions about \bugswarm, and re-affirm
its goals and intended use in \cref{sec:discussion}.

\section{Overview on \bugswarm}
\label{sec:bugswarm}
\begin{figure}[t]
    \center
    \includegraphics[scale=0.45]{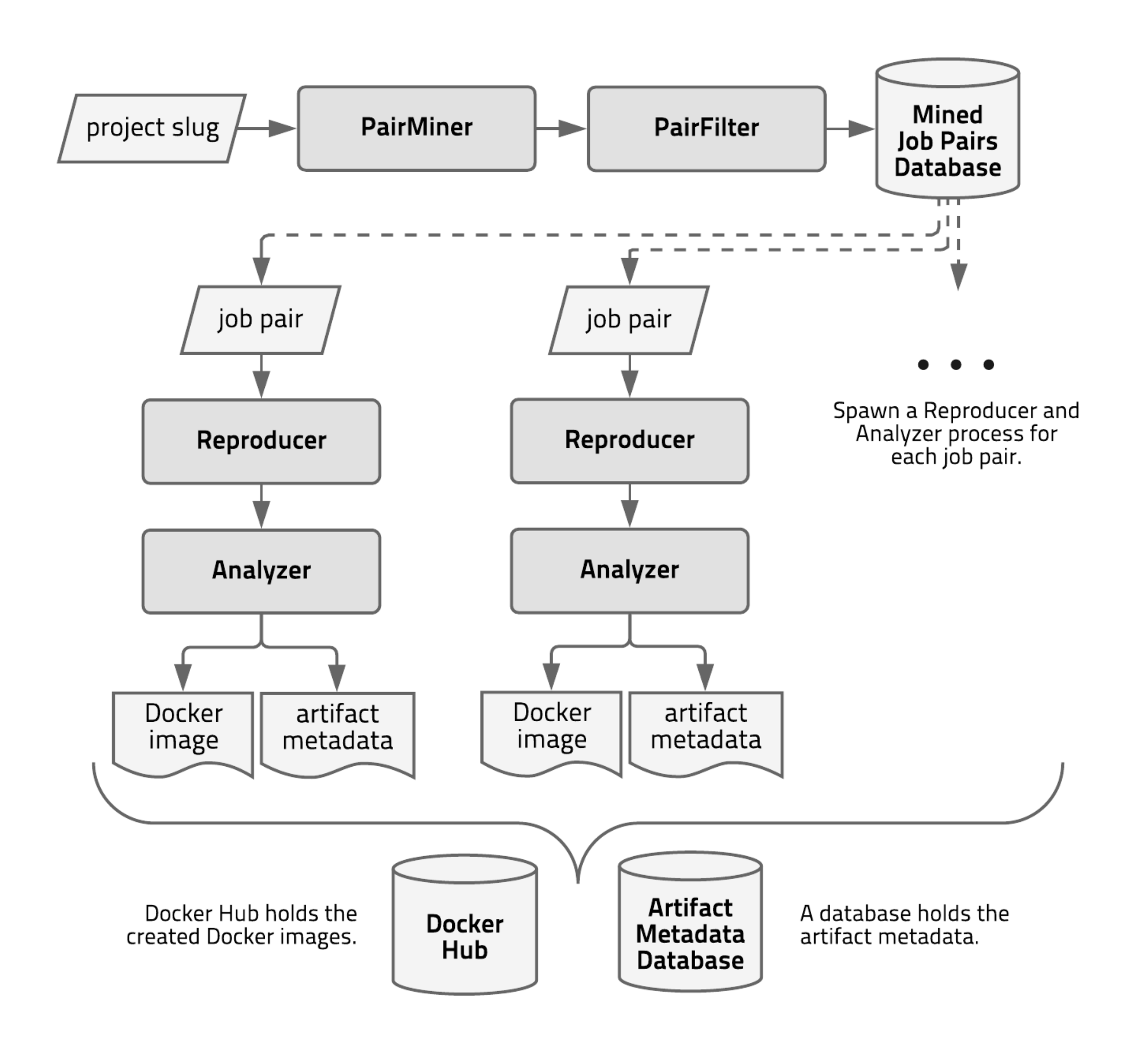}
    \caption{Workflow for the \bugswarm toolkit}
    \label{bugswarm-arch}
\end{figure}

\bugswarm is comprised of three main components: (1) an
infrastructure\footnote{\url{https://github.com/BugSwarm} -- in the
process of open sourcing.} to automatically mine \emph{and} reproduce
failures and fixes from open-source projects that use continuous
integration (\textsc{Travis-CI}), (2) a continuously growing dataset
of real-world failures and fixes packaged in publicly available Docker
images to facilitate
reproducibility,\footnote{\url{https://hub.docker.com/r/bugswarm/images/tags}}
and (3) a website\footnote{\url{http://www.bugswarm.org/dataset/}} and
a REST API\footnote{\url{https://github.com/BugSwarm/common}} for
dataset users to navigate and select artifacts based on a number of
characteristics.

\subsection{\bugswarm Infrastructure}

\bugswarm's methodology to create a continuously growing dataset of
real-world failures and fixes is shown in \cref{bugswarm-arch}. We
briefly describe each component below. For more details please refer
to the \bugswarm paper \citep{DBLP:conf/icse/DmeiriTWBLDVR19}. 

\subsubsection*{\emph{PairMiner.}}
PairMiner represents the first stage of the process. The role of
PairMiner is to mine fail-pass job pairs from the \textsc{Travis-CI}'s
build history of open-source projects hosted in GitHub. A project's
build history refers to all \textsc{Travis-CI} builds previously
triggered. A build may include many jobs; for example, a build for a
Python project might include separate jobs to test with Python
versions 2.6, 2.7, 3.0, etc. The input to PairMiner is the repository
slug (e.g., google/auto) of the project of interest. PairMiner
analyzes the project's build history to identify fail-pass build
pairs, where a build fails and the next consecutive build passes. From
these fail-pass build pairs, PairMiner will extract fail-pass
\emph{job pairs}. The output of PairMiner is a set of fail-pass job
pairs found for the given project.

\subsubsection*{\emph{PairFilter.}}
PairFilter takes as input the \textsc{Travis-CI} fail-pass job pairs
from PairMiner and ensures that essential data is available to allow
for reproduction: (1) the state of the project at the time the job was
executed, and (2) the environment in which the job was executed. If
these essentials are not available then PairFilter will discard the
fail-pass job pair. PairFilter will determine the Docker image that
was the exact build environment for the fail-pass job pair and the
specific commits that triggered each job. The output of PairFilter is
the subset of fail-pass job pairs for which (1) and (2) are available.

\subsubsection*{\emph{Reproducer.}}
The goal of Reproducer is to reproduce each job in the fail-pass job
pair in the same build environment as it was originally run. The input
to Reproducer is a fail-pass job pair, the commits for each version,
and the Docker image for the build environment. Reproducer conducts
the following: (1) generates a job script, i.e., a shell script to
build the project and run regression tests, (2) matches the build
environment, as the job was originally ran in, via a Docker image from
the PairFilter, (3) reverts the project to the specific version, and
(4) runs the code for the job in the Docker image via the job script.
The Reproducer can be ran in parallel via multiple processes for each
job pair as shown in \cref{bugswarm-arch}. The output of Reproducer is
a build log, which is a transcript of everything that occurs at the
command line during the build and testing process. 

\subsubsection*{\emph{Analyzer.}}
The Analyzer parses the original (historical) and reproduced build
logs, extracts key attributes, and compares the extracted attributes
to ensure they match. The key attributes that are parsed are the
status of the build (passed, failed, or errored), and the result of
the test suite (number of tests ran, number tests failed, and names of
failed tests). If the results match between the original and
reproduced build logs, then metadata about the pair will be added to
the \bugswarm database.

\subsubsection*{\emph{Artifact Creation.}}
The Reproducer and Analyzer are run five times. If a fail-pass job
pair was reproducible all five times then we mark it as
``reproducible''. If the number of times the pair was reproducible was
less than five but more than zero then it will be marked as ``flaky''.
A pair can be flaky due to a variety of reasons but primarily because
of test flakiness which can be caused by non-deterministic tests due
to concurrency or environmental changes. Lastly, if a pair is
reproducible zero times then it will be marked as ``unreproducible''.
A reproducible or flaky job pair is referred to as a \bugswarm
artifact.

For each \bugswarm artifact, a Docker image is created which has both
versions of the code and the job scripts to build and test each
version. This Docker image is then stored on our DockerHub
repository.\footnote{\url{https://hub.docker.com/r/bugswarm/images}}
We chose to package each \bugswarm artifact in a Docker image because
Docker facilitates reproducibility. Docker is also a good choice
because it is light-weight, and uses layering. Docker images are
composed of multiple layers which can be shared across multiple Docker
images to save space. Docker does not re-download or store a layer
that is already on a system \citep{docker-storage}.

\subsection{\bugswarm Dataset}
The \bugswarm dataset is the first continuously growing dataset of
reproducible real-world failures and fixes. The dataset was
automatically created using the \bugswarm infrastructure without
controlling for any specific attributes. Currently, the \bugswarm
dataset (version 1.1.0) consists of 3,140 artifacts that are written
in Java and Python. There are a diverse number of artifacts with
different build systems ranging from Maven, Gradle, and Ant to
different longevity from 2015 to 2019 and different testing frameworks
such as JUnit and unittest. We expect a steady grow of the dataset in
the next months as the \bugswarm infrastructure is set to run in
dedicated servers.

\subsection{\bugswarm Website and REST API}
\bugswarm offers many different characteristics to filter by to create
a subset that is useful in the evaluation of a given tool. Examples of
such characteristics are: language, size of diff, build system, number
of tests ran, number of failed tests, patch location (e.g., source
code, test code, or build files), exceptions thrown during run time
(e.g., NullPointerException), etc. The \bugswarm website and REST API
allow the selection of artifacts based on the above attributes.

\section{Methodology by \criticalreview \citep{DBLP:journals/corr/abs-1905-09375}}
\label{sec:critical-review}

The goal of \criticalreview's study is to answer the following
questions:

\begin{rqs}
    \item \label{item:RQ1} What are the main characteristics of
    \bugswarm's pairs of builds regarding the requirements for APR and
    FL?
    \item \label{item:RQ2} What is the execution and storage cost of \bugswarm?
    \item \label{item:RQ3} Which pairs of builds meets the
    requirements of APR and FL?
\end{rqs}

\subsubsection*{\emph{Characteristics of \bugswarm's Pairs of Builds.}}
\criticalreview characterizes the \bugswarm dataset with respect to
requirements of current APR and FL tools: (1) behavioral bugs, (2)
test suite is used with passing tests defining correct behavior and
failing tests defining incorrect behavior, (3) execution set up is
known in terms of path of source, test files, etc., (4) uniqueness of
bugs, and (5) human patch availability. The above requires, for each
artifact, the source code for the buggy version and the fixed version,
the diff between the two versions, and the \textsc{Travis-CI} build
log for the failing job. 

\criticalreview queries for \emph{fully reproducible} Java and Python
artifacts  (see \cref{sec:number-artifacts} for further details) using
the \bugswarm REST API. The resulting artifacts are then filtered for
unique commits.\footnote{Note that multiple
\textsc{Travis-CI} jobs may originate from a single \textsc{Travis-CI}
build.} The diff of each artifact is calculated by retrieving the
buggy and fixed versions of the artifact from its corresponding Docker
image, pushing the code into a branch of a new GitHub repository, and
then invoking the GitHub API to retrieve the diff between the two code
versions. Unique diffs are identified based on md5 hash values, and
artifacts are classified based on whether the extension of the changed
files are \texttt{.java} or \texttt{.py}. Lastly, a high-level
classification of the reason of failure is conducted by using regular
expressions to match certain patterns (test failures, style checkers,
compilation errors, etc.) on \textsc{Travis-CI} build logs.

\subsubsection*{\emph{Execution and Storage of \bugswarm.}}
\criticalreview estimates the size of the \bugswarm dataset for
download and storage, as well as its usage cost. The size of the
dataset is calculated using two metrics: counting \emph{every} Docker
layer, and counting every \emph{unique} Docker layer. Note that Docker
does not download or store a layer that is already in the system (see
\citep{docker-storage} and \cref{sec:execution-storage}).
\criticalreview gives a time estimate for download assuming a 80
Mbit/s stable connection. Finally, the cost of using the full dataset
is estimated assuming a 20-minute experiment per artifact using Amazon
Cloud Instances.

\subsubsection*{\emph{Pairs for APR and FL.}}
\criticalreview lists what the paper considers the requirements to use
state-of-the-art APR and FL tools: (1) artifacts that have been
reproduced five times, (2) artifacts whose Docker images are
available, (3) non empty diff, (4) unique commit, (5) unique diff, (6)
test case failure, and (7) only source files changed. \criticalreview
then reports the number of \bugswarm artifacts that satisfy those
requirements.

\section{Incorrect Claims by \criticalreview \citep{DBLP:journals/corr/abs-1905-09375}}
\label{sec:corrections}

After replicating the study presented by \criticalreview and
inspecting its scripts, we identified incorrect claims made by
\criticalreview related to inconsistencies in the number of artifacts
reported in the \bugswarm paper
\citep{DBLP:conf/icse/DmeiriTWBLDVR19}, a misleading duplication of
commits in the dataset, and calculations of the storage required by
the dataset. Below we discuss each incorrect claim, organized per
research question as presented in
\citep{DBLP:journals/corr/abs-1905-09375}.

\subsection{RQ1: Characteristics of \bugswarm's Pairs of Builds}

\subsubsection*{\emph{Incorrect Number of Artifacts.}} \label{sec:number-artifacts}
\criticalreview reports the number of ``builds'' reproduced five times
given a \bugswarm API request listed in \citep[Section
III-B]{DBLP:journals/corr/abs-1905-09375}.\footnote{http://www.api.bugswarm.org/v1/artifacts/?where=\{``reproduce\_successes'':
\{``\$gt'':4,``lang'':\{``\$in'':[``Java'',``Python'']\}\}\} } The API
request returns 2,949 artifacts while the \bugswarm
paper\citep{DBLP:conf/icse/DmeiriTWBLDVR19} gives 3,091 artifacts.
Thus, \criticalreview reports a contradiction by the \bugswarm
authors, which according to \criticalreview had stated that each
``build'' in the dataset was successfully reproduced five times.

\newpage

\criticalreview states in \citep[Section III-C]{DBLP:journals/corr/abs-1905-09375}:
\begin{displayquote}
Indeed, we considered all pairs of builds that are reproduced
successfully five times like it is described in \bugswarm's paper (see
Section 4-B). Surprisingly, \bugswarm authors did not consider their
criteria in their final selection of the pairs of builds and
consequently the reported number is in contradiction with the paper.
\end{displayquote}

\bugswarm original paper states in \citep[Section IV-B]{DBLP:conf/icse/DmeiriTWBLDVR19}:
\begin{displayquote}
We repeated the reproduction process 5 times for each pair to
determine its stability. If the pair is reproducible all 5 times, then
it is marked as 'reproducible'. If the pair is reproduced only
sometimes, then it is marked as 'flaky'. Otherwise, the pair is said
to be 'unreproducible'.
\end{displayquote}

First, as discussed in the \bugswarm paper \citep[Section
III-C]{DBLP:conf/icse/DmeiriTWBLDVR19} and in \cref{sec:bugswarm} of
this paper, \bugswarm is comprised of artifacts (\textsc{Travis-CI}
job pairs), thus a request from the \bugswarm API will return the
number of \emph{artifacts}, not the number of \emph{builds}. 

Second, the \bugswarm API request used by \criticalreview is returning
the number of artifacts \emph{successfully} reproduced five times. In
other words, the query is returning the number of fully reproducible
artifacts. However, the \bugswarm dataset \citep[Table
III]{DBLP:conf/icse/DmeiriTWBLDVR19} includes both fully reproducible
and flaky artifacts, which together account for a total of 3,091
artifacts. The correct \bugswarm REST API
request\footnote{http://www.api.bugswarm.org/v1/artifacts/?where=\{``reproduce\_successes'':\{``\$gt'':0\},``reproduce\_attempts'':5,``lang'':\{``\$in'':[``Java'',``Python'']\}\}\}}
needs to filter based on a number of reproduce successes greater than
zero and a number of attempts equal to five. All 3,091 artifacts
included in the dataset were attempted five times.

At the time \criticalreview was written (\bugswarm dataset 1.0.1 from
May 2019\footnote{\url{http://www.bugswarm.org/releases/}}), the
number of fully reproducible artifacts was indeed 2,949 and the number
of flaky artifacts was 142. There is no contradiction on the selection
criteria described in the \bugswarm paper: both reproducible and flaky
artifacts are included in the dataset.

\subsubsection*{\emph{Duplicate Failing Commits.}} \criticalreview
reports a ``new'' finding regarding a high number of duplicate failing
commits in the \bugswarm dataset that would introduce misleading
results.

\criticalreview states in \citep[Section
II-C]{DBLP:journals/corr/abs-1905-09375}:
\begin{displayquote}
Our second observation is that 40.08\% ((2,949-1,767)/2,949) of the
builds have a duplicate failing commit. It means that those 40.08\%
should not be considered by the approaches that only consider the
source code of the application otherwise it introduces misleading
results.
\end{displayquote}

\newpage

\bugswarm paper states in \citep[Section IV-B]{DBLP:conf/icse/DmeiriTWBLDVR19}:
\begin{displayquote}
Recall from Section III-C that \textsc{PairMiner} mines job pairs. The
corresponding number of reproducible \emph{unique} build pairs is
1,837. The rest of the paper describes the results in terms of number
of job pairs.
\end{displayquote}

As stated in the \bugswarm paper \citep[Section
III-C]{DBLP:conf/icse/DmeiriTWBLDVR19} and in \cref{sec:bugswarm} of
this paper, a \bugswarm artifact corresponds to a pair of \emph{jobs},
not a pair of \emph{builds} (as incorrectly interpreted throughout
\criticalreview). A \textsc{Travis-CI} build can be composed of
multiple jobs that test the same commit under different
configurations. Early feedback from researchers in our community
indicated that such artifacts can also be of interest to researchers.

As also described in the \bugswarm paper \citep[Section
III-B]{DBLP:conf/icse/DmeiriTWBLDVR19}, a given experiment may require
artifacts that meet specific criteria. If such criteria require
uniqueness of job pairs, as reported by \criticalreview is the case
for APR tools, then we provide a REST API and website that allow to
consider uniqueness when selecting artifacts of interest. Thus, having
the dataset include multiple jobs from a build does not represent a
problem that would introduce misleading results.\footnote{The
difference between 1,767 and 1,837 is again due to \criticalreview
omitting flaky artifacts.}

\subsection{RQ2: \bugswarm Execution and Storage Cost} \label{sec:execution-storage}
\begin{table}[t!]
    \caption{Table of Metrics of \bugswarm Downloading and Storage Cost
    from \citep{DBLP:journals/corr/abs-1905-09375}.}
    \begin{center}
        \begin{tabular}{| l|r|r|r |}
        \hline
        Metrics in Gigabytes (GB) & Java & Python & All\\
        \hline \hline
        \bugswarm Docker layer size & 5,107 & 3,813 & 8,921\\
        \bugswarm \textbf{unique} Docker layer size & \textbf{1,327} & \textbf{919} & \textbf{2,246}\\
        Avg. size & 3.01 & 3.05 & 3.03 \\
        Download all layers (80Mbits/s) & 6d, 7.8h & 4d, 17.13h & 11d, 1.16h\\
        Download \textbf{unique} layers (80Mbits/s) & \textbf{1d,
        15.4h} & \textbf{1d, 3.3h} & \textbf{2d, 18.8h}\\
        \hline
        \end{tabular}
    \end{center}
    \label{tab:storage}
\end{table}

\criticalreview calculates the size of the \bugswarm dataset and
provides estimated download time and cost for using the full dataset
in Amazon Web Instances \citep[Section
3-D]{DBLP:journals/corr/abs-1905-09375}. The paper reports that the
full dataset is 8,921 GB, which takes about 11 days 1.16 hours to
download when using a 80 Mbits/s internet connection. Subsequently,
the cost of using the \bugswarm dataset, assuming a 20-minute
experiment, is \$711.30 USD.

\subsubsection*{\emph{Download Size Calculation.}}
\criticalreview calculates the size of the \bugswarm dataset using two
metrics: counting \emph{every} Docker layer, and counting every
\emph{unique} Docker layer. The size of the dataset is reported (see
\cref{tab:storage} from \criticalreview) as 8,921 GB and 2,246 GB,
respectively. However, counting every Docker layer is incorrect.
Docker does \emph{not} re-download or store a layer that is already on
a system \citep{docker-storage}. The average size (row 3) and download
time (row 4) given in \cref{tab:storage} are calculated based on
\emph{all} Docker layers (row 1), thus these table entries are also
incorrect.

\subsubsection*{\emph{Compression Ratio.}} \criticalreview estimates a
compression ratio used to incorrectly calculate space in disk. A
compression ratio is unnecessary in the first place; disk space is
determined by the size of unique Docker layers, already given in row 2
of \cref{tab:storage}.

\criticalreview states in \citep[Section III-D]{DBLP:journals/corr/abs-1905-09375}: 

\begin{displayquote}
According to our observations, the ratio between download size and
disk storage is 2.48x and drops to 0.41x when considering the
duplicate layers. [...] Based on this observation, we estimate the
total disk space required to 3,680.45 GB.
\end{displayquote}

\criticalreview fails to mention that the above observations are based
on 464 artifacts \citep{downloaded-images}, not the full dataset. The
script \citep{space-on-disk} used to calculate disk space lists 598.98
GB of storage used by the 464 artifacts. When we downloaded the same
464 artifacts, the disk space reported by the command \texttt{docker
system df} is 353 GB, not 598.98 GB. 

The compression ratio is then calculated by dividing the space in disk
by the size of the 464 artifacts when considering all Docker layers:
598.98 GB / 1,452.02 GB = 0.42. However, when using this compression
ratio, the estimated disk space reported for the full dataset is
3,680.45 GB, which is 63\% higher than the actual size given in
\cref{tab:storage} in row 2, which is 2,246 GB.

\subsubsection*{\emph{Cost Calculation.}} Because the cost of using
the \bugswarm dataset is based on incorrect estimated download and
storage sizes, the cost calculations are also incorrect. Additionally,
as mentioned earlier and corroborated by \criticalreview, we expect
that \bugswarm users will be interested in subsets of the dataset, as
opposed to the full dataset. This must be taken into account when
making such cost calculations.

\section{Other Contributions \& Findings by \criticalreview\citep{DBLP:journals/corr/abs-1905-09375}}
\label{sec:other-contributions}

In addition to answering the questions described in
\cref{sec:critical-review}, \criticalreview also provides a GitHub
repository for the \bugswarm artifacts and a website to navigate and
select artifacts. This section discusses these contributions as well
as a finding regarding duplicate diffs.

\subsubsection*{\emph{GitHub Repository.}}
One of the contributions listed by \criticalreview is a new GitHub
repository\footnote{\url{https://github.com/TQRG/BugSwarm}} to store
\bugswarm artifacts. Specifically, there is a branch for each artifact
that contains the buggy version of the code, the fixed version of the
code, the diff between both versions, and the failing and passing
\textsc{Travis-CI} build logs. The only artifact information not
stored in the repository is the scripts to build the code and run
regression tests.

However, the existence of the \criticalreview repository is \emph{not}
necessary. The buggy and fixed versions of the code can be directly
accessed via the original repositories (the \bugswarm REST API and the
website provide the commit information) or by downloading the
\bugswarm Docker image for the artifact, which includes a copy of both
versions of the code. The \textsc{Travis-CI} build logs can be
directly accessed via the \textsc{Travis-CI} website using the
information provided by the \bugswarm REST API or directly following
the \bugswarm website links. Finally, the diff can be directly
retrieved using the GitHub API (3-dot diff), or accessed via the
\bugswarm website (2-dot diff).

\subsubsection*{\emph{Website to Browse and Select \bugswarm Artifacts.}}
Another contribution listed by \criticalreview is a
website\footnote{\url{https://tqrg.github.io/BugSwarm}} to browse and
select \bugswarm artifacts. The website displays the number of
added/removed/modified lines and files, and allows to select artifacts
based on unique commits, unique diffs, not empty diffs, containing
failing tests, changing source code, a manual categorization of
bug/non-bug patches, and a high-level categorization of failures.

\bugswarm
already provides its own
website\footnote{\url{http://www.bugswarm.org/dataset/}} for browsing
and selection based on the same attributes listed by \criticalreview
(except for their two categorizations, which are complementary to our
own). The \bugswarm website also allows to select artifacts based on
the location of the fix (source files, configuration files, or test
files). In addition to the website, \bugswarm provides a REST API to
query the \bugswarm database directly, thus one is not restricted to
the options provided in the website. \bugswarm provides a
classification of artifacts based on runtime exceptions.

\subsubsection*{\emph{Duplicate Diffs.}}
\criticalreview reports that while controlling for unique failed
commits there are duplicate diffs among them, reporting that 198 out
of 1,767 artifacts have a duplicate diff \citep[Section
III-C]{DBLP:journals/corr/abs-1905-09375}. 

Recently, we have discovered that \textsc{Travis-CI} can make a
``double build'' when a build is a Pull Request
(PR).\footnote{\url{https://docs.travis-ci.com/user/pull-requests/\#double-builds-on-pull-requests}}
\textsc{Travis-CI} will create a build for the PR branch, and another
build for the PR branch merged with the base branch. If no changes
have been made to the base branch since the time the PR branch was
created, then the diffs between both builds will be the same. This
explains \criticalreview's observation. Fortunately, we believe it is
feasible to automatically detect these cases, and this detection will
be incorporated into the \bugswarm infrastructure to avoid such cases
in future versions of the dataset.

\section{Discussion on the Role of \bugswarm}
\label{sec:discussion}

We would like to conclude this paper by briefly clarifying a few
misconceptions about \bugswarm, and by discussing our vision for the
\bugswarm infrastructure and dataset.

\textbf{(1) \bugswarm is more than a dataset}. As described in
\cref{sec:bugswarm}, \bugswarm is comprised of an infrastructure to
automatically create a large-scale dataset of real-world failures and
fixes, a continuously growing dataset, and a REST API and website to
navigate and select artifacts from the dataset based on
characteristics of interest.

\textbf{(2) The \bugswarm dataset is not static.}  One of the main
contributions of the \bugswarm infrastructure is that its full
automation has enabled the creation of a \emph{continuously} growing
dataset. As discussed in the \bugswarm paper
\citep{DBLP:conf/icse/DmeiriTWBLDVR19}, the potential for size and
diversity opens new opportunities, but it also presents several
challenges. Some of these challenges include data versioning
(discussed in \criticalreview), and automated bug classification to
increase the usefulness of the dataset.

\textbf{(3) The \bugswarm dataset is not meant for a single target
application.} Because of the size and diversity of the \bugswarm
dataset, it is unrealistic to believe that all artifacts will be
relevant to one application. As a result, \bugswarm facilitates
navigating and selecting artifacts based on a set of characteristics
via the \bugswarm website or REST API. Thus, it is easy to select
artifacts for a given application (e.g., APR or FL) beforehand.

\textbf{(4) \bugswarm artifacts with specific characteristics can be
``grown''.} The initial \bugswarm dataset was created without
controlling for any particular attribute, such as diff size, patch
location, or reason for failure. However, since the publication of the
\bugswarm paper \citep{DBLP:conf/icse/DmeiriTWBLDVR19}, target mining
is now available and thus, it is possible to grow the dataset in
specific directions. We believe that allowing for diverse
characteristics does not hinder the evaluation of the state of the
art. On the other hand, we hope that the existence of artifacts that
the state of the art may not be able to handle today will further push
advancement.

\bugswarm is a project under active development, and in process of
open sourcing its infrastructure. We welcome feedback from the
community. The \bugswarm dataset is publicly available in DockerHub.
The website is also publicly available, and the REST API is available
to anyone who would like to request a token to access the \bugswarm
database.

\section*{Acknowledgments}

We thank Bohan Xiao and Octavio Corona for their help in gathering
some of the data discussed in this paper. This work was supported by
NSF grant CNS-1629976 and a Microsoft Azure Award. Any opinions,
findings, and conclusions or recommendations expressed in this
material are those of the authors and do not necessarily reflect the
views of the National Science Foundation or Microsoft.

\bibliographystyle{splncs04nat}
\bibliography{main}

\end{document}